\documentclass{emulateapj}

\usepackage{graphicx}
\usepackage{psfig}
\usepackage{apjfonts}

\def\cm{\textrm{cm}}

\def\erg{\textrm{erg}}

\def\Kelv{\textrm{K}}

\def\sr{\textrm{sr}}

\def\gcm2{\textrm{g}~\textrm{cm}^{-2}}

\def\phcm2s1{\textrm{photons}~\textrm{cm}^{-2}~\textrm{s}^{-1}}

\def\eV{\textrm{eV}}
\def\keV{\textrm{keV}}
\def\MeV{\textrm{MeV}}
\def\GeV{\textrm{GeV}}

\def\MHz{\textrm{MHz}}
\def\GHz{\textrm{GHz}}

\def\muGauss{\mu\textrm{G}}
\def\mGauss{\textrm{mG}}

\newcommand{\mean}[1]{\ensuremath{\langle #1 \rangle}}

\begin{document}

\title{The $\gamma$-Ray Background Constrains the Origins of the Radio and X-Ray Backgrounds}
\author{Brian C. Lacki\altaffilmark{1,2}}
\altaffiltext{1}{Department of Astronomy, The Ohio State University, 140 West 18th Avenue, Columbus, OH 43210, USA, lacki@astronomy.ohio-state.edu}
\altaffiltext{2}{Center for Cosmology \& Astro-Particle Physics, The Ohio State University, Columbus, Ohio 43210, USA}

\begin{abstract}
Cosmic ray protons generate $\gamma$-rays, neutrinos, and secondary electrons and positrons ($e^{\pm}$) through pion-producing collisions with gas atoms.  Any synchrotron or Inverse Compton (IC) radiation from secondary $e^{\pm}$ is therefore accompanied by pionic $\gamma$-rays.  Using the extragalactic $\gamma$-ray background, we constrain the contribution of secondary $e^{\pm}$ to the cosmic radio, X-ray, and soft $\gamma$-ray backgrounds.  These bounds depend on the unknown hadronic contribution to the $\gamma$-ray background and the backgrounds' source redshifts.  With our assumptions, we find that IC-upscattered light from secondaries is $\la 1/4$ of the MeV - GeV $\gamma$-ray background and $\la 10\%$ of the 0.5 keV - 1 MeV background (for sources at a redshift $z \la 10$).  The low intensity of the observed $\gamma$-ray background is marginally inconsistent with a secondary $e^{\pm}$ origin for the radio background reported by ARCADE at $\sim 3\ \GHz$, unless the magnetic field strength in their sources is milliGauss or greater.  These limits on the magnetic field strength are sensitive to uncertainties.  However, any contribution to the $\gamma$-ray background from sources not responsible for the ARCADE excess increases the inconsistency.
\end{abstract}


\section{Introduction}
\label{sec:Introduction}
Cosmic rays (CRs) are accelerated in many environments including star-forming galaxies \citep[SFGs; e.g.,][]{Condon92} and galaxy clusters \citep[e.g.,][]{Ferrari08,Rephaeli08}.  The bulk of the CR energy is in protons.  These collide with ambient nuclei, creating pions, which decay into $\gamma$-rays, neutrinos, and secondary electrons and positrons ($e^{\pm}$).  Whether secondary or primary, CR $e^{\pm}$ radiate synchrotron emission in magnetic fields and Inverse Compton (IC) as they scatter low energy photons.  CR protons therefore contribute to the $\gamma$-ray and neutrino backgrounds, while CR $e^{\pm}$ contribute to the radio, X-ray, and $\gamma$-ray backgrounds.

The origins of these backgrounds are understood to varying degrees.  The $\gamma$-ray background was once attributed to blazars, but \emph{Fermi} has revealed that another source may be responsible for most of the emission above 100 MeV \citep{Abdo10a}.  SFGs are one explanation for the $\gamma$-ray background \citep[e.g.,][]{Fields10,Lacki10a}.  The neutrino background is not yet detected, although IceCube will improve sensitivity greatly \citep{Achterberg07}.  The radio background is assumed to come from CR $e^{\pm}$ in SFGs and possibly AGNs \citep{Protheroe96,Haarsma98,Dwek02}.  However, the radio bolometer ARCADE detected an extragalactic radio background six times greater than expected from the radio luminosities of $z \approx 0$ galaxies \citep{Fixsen09,Seiffert09}.  \citet{Singal09} suggested that redshift evolution of the radio properties of SFGs explains the ARCADE background.  The X-ray background is the best understood, with most of it being resolved into AGNs \citep[][and references therein]{Gilli07}.  

A powerful way of limiting one cosmic background is to compare it with another of the same origin.  For example, the Waxman-Bahcall argument limits the flux of ultra-high energy neutrinos from the observed spectrum of ultra-high energy CR protons which produce the neutrinos \citep{Waxman99,Bahcall01}.  Simply put, the Universe must be at least as luminous in the protons that generate secondaries as in the secondaries themselves.  Similarly, we can use one pionic background -- either the $\gamma$-rays or neutrinos -- to constrain the others: synchrotron radio or IC X-rays from secondary $e^{\pm}$.  Secondary $e^{\pm}$ may dominate over primary electrons in starburst galaxies \citep[e.g.,][]{Thompson07} and possibly galaxy clusters \citep[e.g.,][]{Dennison80}, so this argument applies to backgrounds from these objects. 

\section{Ratio of Pionic $\gamma$ Rays to Secondary Emission}
Suppose a class of sources emits CR protons of energy $E_p$, which experience pionic losses during their propagation.  The pions decay, generating $\gamma$-ray, neutrino, and leptonic backgrounds of flux intensity $dJ/dE$, so the power in each of the backgrounds per log bin energy is $E dJ/dE$, where $E$ is the energy of the decay product.  About 1/3 of the energy lost to pionic interactions goes into neutral pions, which decay into $\gamma$-rays with typical energy $\mean{E_{\gamma}} \approx 0.1 E_p$.  The remaining energy is in charged pions; of this, 1/4 goes into secondary $e^{\pm}$ and the rest into neutrinos, so 1/6 of the pionic luminosity is in secondary $e^{\pm}$ while 1/2 is in neutrinos.  The average energy of the neutrinos and $e^{\pm}$ is $\mean{E_e} \approx \mean{E_{n}} \approx 0.05 E_p \approx \mean{E_{\gamma}} / 2$.  Taking the ratio of the background intensity in pionic $\gamma$-rays to pionic secondary $e^{\pm}$, we have:
\begin{equation}
2 \mean{E_e} \frac{dJ_e}{dE_e} \approx \mean{E_{\gamma}} \frac{dJ_{\gamma}}{dE_{\gamma}},
\end{equation}
and similarly, $3 \mean{E_e} \frac{dJ_e}{dE_e} \approx \mean{E_{n}} \frac{dJ_{n}}{dE_{n}}$ for neutrinos.  In comparing $\mean{E_{\gamma}}$ to $\mean{E_e}$, we assume the pions are relativistic; we take $E_{\gamma} \ge 0.3 E_{0.3}\ \GeV$ as a threshold for this.  Far below this energy, few secondaries are expected and any emission comes from primary $e^{\pm}$.

The secondary $e^{\pm}$ radiate synchrotron and IC emission, among other losses.  The pitch-angle averaged rest-frame frequency of synchrotron emission is $\nu_C = (3 E_e^2 e B)/(16 m_e^3 c^5)$, where $e$ is the electron's charge and $B$ is magnetic field strength.  Since $\nu_C \propto E_e^2$, $d\ln\nu_C = 2 d\ln E_e$: the synchrotron emission from one log bin in $e^{\pm}$ energy is spread over two log bins in synchrotron frequency.  At most $100\%$ of the CR $e^{\pm}$ emission can go into synchrotron, implying that $\nu_C dJ_{e}/d\nu_C (\nu_C) = (E_e/2) dJ_e / dE_e$, or 
\begin{equation}
\label{eqn:DiffSynchLimit}
\nu_C \frac{dJ_{e}}{d\nu_C} ({\rm synch}) \la (f/4) E_{\gamma} \frac{dJ_{\gamma}}{dE_{\gamma}} ({\rm pionic~\gamma-ray})
\end{equation}
evaluated for $\nu_C$ at $E_{e} = E_{\gamma} / 2$, where $f \approx 1$ parameterizes uncertainties in this approximation and the backgrounds \citep{Loeb06}.  Lower $f$ linearly scales down the $\gamma$-ray background, either because the total background is lower than assumed here or to consider only the pionic contribution from some class of sources; similarly, higher $f$ linearly corresponds to lower synchrotron or IC backgrounds, either from errors in the measured backgrounds, or to consider only the contribution from secondaries from some source class.  This uses the $\delta$-function approximation for the synchrotron spectrum, which is generally valid for power law spectra \citep[e.g.,][]{Felten66}.\footnote{This approximation is accurate to $\sim 25\%$ for an $E^{-2}$ steady-state $e^{\pm}$ spectrum and is even better for an $E^{-3}$ spectrum.  Note that $70\%$ of the synchrotron emission of electrons with $E_e$ is in the 2 ln bins centered on $\nu_C$.} 

Similarly, the average rest-frame energy of an IC upscattered photon of initial energy $\epsilon_0$ is $E_{\rm IC} = (4 E_e^2 \epsilon_0)/(3 m_e^2 c^4)$ in the Thomson limit ($E_{\rm IC} \la E_e$).  Once again $E_{\rm IC} \propto E_e^2$, and the IC emission from one log bin in $E_e$ is spread over two log bins in $E_{\rm IC}$.  We have
\begin{equation}
\label{eqn:DiffICLimit}
\nu_{\rm IC} \frac{dJ_{e}}{d\nu_{\rm IC}} ({\rm IC}) \la (f/4) E_{\gamma} \frac{dJ_{\gamma}}{dE_{\gamma}} ({\rm pionic~\gamma-ray})
\end{equation}
evaluated for $E_{e} = E_{\gamma} / 2$, again using the $\delta$-function approximation \citep{Felten66}.  Note that eqs.~\ref{eqn:DiffSynchLimit} and \ref{eqn:DiffICLimit} apply not just to the whole backgrounds, but to the pionic emission from \emph{each} source and \emph{each} population.

In what follows, we conservatively assume that \emph{all} of the observed $\gamma$-ray background \citep{Abdo10b} is pionic in origin.  Removing leptonic contributions to the $\gamma$-ray background only tighten the limits.  The power-law fit to the \citet{Abdo10b} background is:
\begin{equation}
\label{eqn:FermiFit}
E_{\gamma} \frac{dJ_{\gamma}}{dE_{\gamma}} ({\rm \gamma-ray}) = 2.33 \times 10^{-9} \left(\frac{E_{\gamma}}{100\ \MeV}\right)^{-0.41}
\end{equation}
in cgs units of $\erg\ \cm^{-2}\ \sec^{-1}\ \sr^{-1}$.  At energies below \emph{Fermi} observations, the observed total $\gamma$-ray background is bounded by eq.~\ref{eqn:FermiFit} (\citealt{Weidenspointner00,Strong04}; see Fig.~\ref{fig:DiffGammaLimitsIC}); therefore using the observed total $\gamma$-ray background instead of eq.~\ref{eqn:FermiFit} gives even stronger limits on backgrounds from secondary $e^{\pm}$ than found here.  At high energies, eqn.~\ref{eqn:FermiFit} applies only if the Universe is transparent to $\gamma$-rays.  This is correct below 20 GeV, our maximum $E_{\gamma}$ for limits on $z = 10$ sources, and below 100 GeV out to $z \approx 1$  \citep[e.g.,][]{Gilmore09,Finke10}.

\section{The X-ray and Soft $\gamma$-ray Backgrounds}
Nonthermal emission in X-rays has been observed in galaxy clusters, and might be IC-upscattered CMB photons \citep[see the review by][]{Rephaeli08}.  \citet{Moran99} suggested IC upscattered ambient far-infrared (FIR) starlight in starburst galaxies contributes significantly ($\sim 5-10\%$) to the X-ray background.  Since pionic $\gamma$-rays accompany pionic secondary $e^{\pm}$ production, the observed $\gamma$-ray background limits the contribution of secondary $e^{\pm}$ in these sources to the X-ray background.

In the observer-frame, and assuming a typical energy of $3 k T_{\rm CMB} (z)$ for CMB photons, the typical energy of upscattered CMB photons is $E_{\rm IC} \approx E_{\gamma}^2 k [T_{\rm CMB} (0)] (1 + z)^2 / (m_e^2 c^4)$.  Plugging eq.~\ref{eqn:FermiFit} into eq. \ref{eqn:DiffICLimit}, we get:
\begin{equation}
\nu_{\rm IC} \frac{dJ_e}{d\nu_{\rm IC}} \la 2.2 \times 10^{-10} f \left(\frac{E_{\rm IC}}{\keV}\right)^{-0.205} (1 + z)^{0.41},
\end{equation}
in cgs units.  For our assumptions about pion kinematics to be valid, we impose the constraint that $E_{\gamma} \ga 0.3 E_{0.3} (1 + z)^{-1}\ \GeV$: 
\begin{equation}
E_{\rm IC} \ga 81 E_{0.3}^2\ \eV.
\end{equation}
Since the $\gamma$-ray background is only observed for $E_{\gamma} \le 100 E_{100} \GeV$ \citet{Abdo10b}, we also require:
\begin{equation}
E_{\rm IC} \la 9.0 E_{100}^2 (1 + z)^2\ \MeV,
\end{equation}
where $E_{100} \to 0.2$ at high $z$ because the Universe is opaque at energies above 20 GeV.

\begin{figure}
\centerline{\includegraphics[width=8cm]{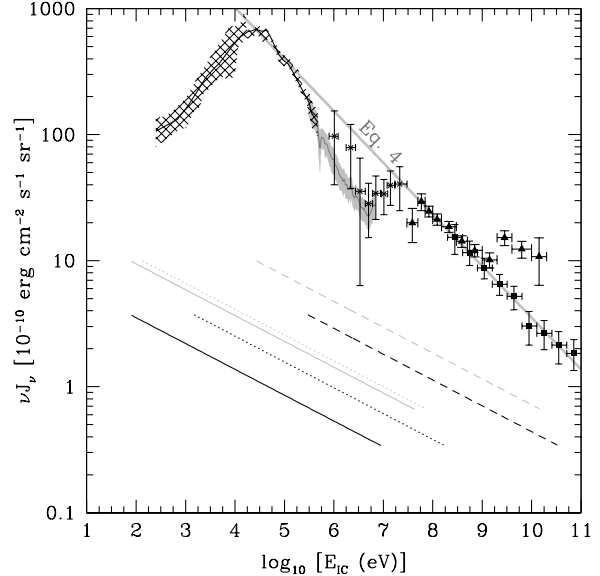}}
\figcaption[figure]{Limits on the X-ray and $\gamma$-ray backgrounds from IC upscattering ($f = 1$, $E_{0.3} = 1$) by secondary $e^{\pm}$ on CMB photons (solid), $50\ \Kelv$ FIR photons (dotted), and $10^4\ \Kelv$ UV/optical photons (dashed), based on the observed $\gamma$-ray background.  Black is $z = 0$ while grey is $z = 10$.  The observed backgrounds are from \citet[][black line and cross-hatching]{Gilli07} and references therein, \citet[grey shading and line]{Watanabe99}, \citet[][Xs]{Weidenspointner00}, \citet[][triangles]{Strong04}, and \citet[][squares]{Abdo10b}.\label{fig:DiffGammaLimitsIC}}
\end{figure}

We proceed similarly for ambient light of temperature $T_{\rm amb}$, finding $E_{\rm IC} \approx E_{\gamma}^2 k T_{\rm amb} (1 + z) / (m_e^2 c^4)$.  Applying eq.~\ref{eqn:DiffICLimit} to eq.~\ref{eqn:FermiFit} gives us in cgs units:
\begin{equation}
\nu_{\rm IC} \frac{dJ_e}{d\nu_{\rm IC}} \la 9.8 \times 10^{-11} f \left(\frac{E_{\rm IC}}{\MeV}\right)^{-0.205} \left(\frac{T_{\rm amb}}{50\ \Kelv}\right)^{0.205} (1 + z)^{0.205},
\end{equation}
valid for $1.5\ \keV E_{0.3}^2 (1 + z)^{-1} T_{50} \la E_{\rm IC} \la 165\ \MeV E_{100}^2 (1 + z) T_{50}$, where $T_{50} = T_{\rm amb} / (50\ \Kelv)$.  

Figure~\ref{fig:DiffGammaLimitsIC} shows that IC-upscattered CMB light from secondary $e^{\pm}$ is only a small fraction of the X-ray background, with greater contributions possible for sources at greater $z$.  For $f = 1$ and sources at $z \approx 0$ (10), it makes up $\la 3\%$ ($\la 7\%$) of the background below 0.5 keV, $\la 1\%$ ($\la 4\%$) at 1 keV, $\la 0.3\%$ ($\la 0.7\%$) at 10 keV, $\la 0.9\%$ ($\la 2\%$) at 1 MeV, and $\la 1\%$ ($\la 3\%$) at 10 MeV.  

As seen in Figure~\ref{fig:DiffGammaLimitsIC}, the bounds on the contribution of upscattered FIR light from secondary $e^{\pm}$ to the X-ray and $\gamma$-ray backgrounds are relatively small.  IC upscattered FIR is $\sim 4 f\%$ or less of the cosmic backgrounds from 1 keV to 1 MeV, and up to $\sim 5 f\%$ of the 1 - 100 MeV background.  Bounds on upscattered optical/UV light from young stars ($T_{\rm amb} = 10000\ \Kelv$) follow similarly.  We find that such emission from secondary $e^{\pm}$ is $\la 16 f\%$ of the actual $\gamma$-ray background for $z = 0$ sources ($\la 9 f\%$ from 1 - 100 MeV), but up to $8 - 15 f\%$ of the 1 - 100 MeV background and up to $\sim f/4$ of the 0.1 - 10 GeV background for sources at $z = 10$.

These results imply that IC emission from secondary $e^{\pm}$ does not contribute significantly to the X-ray or soft $\gamma$-ray backgrounds.  However, they do not apply to primary electrons or to secondary $e^{\pm}$ that have been reaccelerated.

\section{The Radio Background}
SFGs are expected to be a major source of the radio background.  Many estimates of the cosmic radio background (such as \citealt{Protheroe96,Haarsma98,Dwek02}) use the FIR-radio correlation (FRC), a tight linear relation between the FIR and GHz synchrotron luminosities of SFGs \citep[e.g.,][]{Helou85,Condon92,Yun01}.  Recent measurements by ARCADE suggest that the radio background is 6 times larger than expected from applying the FRC to the IR background \citep{Fixsen09,Seiffert09}.  One way to explain this excess is if the FRC evolves with $z$ \citep{Singal09}.  However, most bright galaxies out to $z \approx 2$ seem to lie on the FRC \citep[e.g.,][]{Appleton04,Sargent10}, or show only moderate deviations \citep[e.g.,][]{Ivison10}.  

Recent work by \citet{Lacki10b}, supported by $\gamma$-ray detections of nearby starburst galaxies \citep{Acciari09,Acero09,Abdo10c}, suggests that a conspiracy enforces the FRC in starburst galaxies: secondary $e^{\pm}$ dominate the primary electrons, increasing the radio emission $\sim 10$ times when combined with spectral effects; while bremsstrahlung, ionization, and IC losses suppress the radio emission by a similar factor at 1 GHz \citep[see also][]{Lacki10a}.  An unbalanced conspiracy could enhance radio emission from starbursts \citep{Lacki10c}, but such ``extra'' radio emission comes from pionic secondary $e^{\pm}$, which are accompanied by pionic $\gamma$-rays.  The pionic $\gamma$-ray background sets a hard limit on the synchrotron background from pionic $e^{\pm}$.

Based on the FRC, \citet{Loeb06} and \citet{Thompson07} calculated starbursts' contribution to the neutrino and $\gamma$-ray backgrounds.  Eq.~\ref{eqn:DiffSynchLimit} \emph{inverts} these arguments: the $\gamma$-ray background sets upper limits on the radio background from starbursts.  These limits apply to other sources of the radio background when secondary $e^{\pm}$ dominate their radio emission.

If pion production creates secondary $e^{\pm}$ with source-frame energy $E_e^{\prime}$ radiating synchrotron at observer-frame frequency $\nu_C$, it also creates pionic $\gamma$-rays with source-frame energy $E_{\gamma}^{\prime} \approx 2 E_e^{\prime}$.  The observed $\gamma$-ray background at $E_{\gamma} = E_{\gamma}^{\prime} (1 + z)^{-1}$ therefore limits the synchrotron background from secondary $e^{\pm}$ at
\begin{equation}
\label{eqn:nuCObs}
\nu_C \approx 3.2 \left(\frac{E_{\gamma}}{\GeV}\right)^2 \tilde{B}_{\rm \mu G} \MHz,
\end{equation}
where $\tilde{B}_{\rm \mu G} = (B / \muGauss) (1 + z)$.  

\begin{figure}
\centerline{\includegraphics[width=8cm]{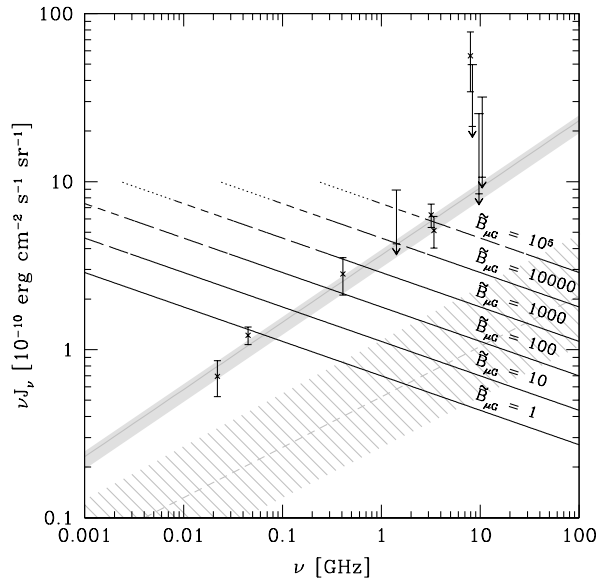}}
\figcaption[figure]{Limits on the radio background ($f = 1$, $E_{0.3} = 1$) from secondary $e^{\pm}$ at a single redshift.  The constraints apply for source populations at $z = 0$ (\emph{solid}), $z = 2$ (\emph{long-dashed}), $z = 5$ (\emph{short-dashed}), and $z = 10$ (\emph{dotted}).  The ARCADE fit is the solid grey line, with uncertainties represented by shading.  The predicted radio background from \citet{Dwek02} (for a $J_{\nu} \propto \nu^{-0.7}$ spectrum) is the dashed grey line, with uncertainies represented by the striped grey area.\label{fig:DiffGammaLimits}}
\end{figure}

The ARCADE fit to the radio background in cgs units is
\begin{equation}
\label{eqn:ARCADE}
\nu_C \frac{dJ_e}{d\nu_C} = 3.7 \times 10^{-10} \nu_{\rm GHz}^{0.4}.
\end{equation}
where $\nu_{\rm GHz}$ is the observed frequency (assumed to be $\nu_C$) in GHz \citep{Fixsen09}.
The errors in the ARCADE data indicate that eq.~\ref{eqn:ARCADE} applies below $\nu_{\rm max} = 3.4\ \GHz$; at higher frequencies, the errors become too large to be sure whether the background spectrum steepens.  If the background is entirely from secondaries, equations \ref{eqn:DiffSynchLimit}, \ref{eqn:FermiFit}, and \ref{eqn:nuCObs} limit the radio background to
\begin{equation}
\label{eqn:FermiSynchLimit}
\nu_C \frac{dJ_e}{d\nu_C} \la 7.0 \times 10^{-11} \nu_{\rm GHz}^{-0.21} \tilde{B}_{\rm \mu G}^{0.21},
\end{equation}
as plotted in Figure~\ref{fig:DiffGammaLimits}.  We obtain a lower limit on $B$ by plugging the ARCADE background (eq.~\ref{eqn:ARCADE}) into eqn.~\ref{eqn:FermiSynchLimit}:
\begin{equation}
\label{eqn:BFromObs}
3.4 (1 + z)^{-1} f^{-4.9} \nu_{\rm GHz}^{2.95} \mGauss \la B.
\end{equation}
At low frequencies, the ARCADE data is easily consistent with the $\gamma$-ray background (below the limits for all $\tilde{B}_{\rm \mu G}$ in Figure~\ref{fig:DiffGammaLimits}).  At higher frequencies, large $\tilde{B}_{\rm \mu G}$ are required: with higher $B$, lower energy $e^{\pm}$ are responsible for the emission at a given frequency, and eq.~\ref{eqn:FermiFit} allows more power at lower $e^{\pm}$ energies.  The limits on $J_e$ are constant in electron energy, but slowly shift in frequency (horizontally in Fig.~\ref{fig:DiffGammaLimits}) with different $B$.

Our limit only applies if $E_{\gamma}^{\prime} \ge 0.3 E_{0.3}\ \GeV$.  This combined with eq.~\ref{eqn:nuCObs} implies that eq.~\ref{eqn:BFromObs} is only valid for $\nu_{\rm GHz} \la 1.01 (1 + z)^{1.03} f^{2.50} E_{0.3}^{-1.03}$ and 
\begin{equation}
\label{eqn:BLim}
B_{\rm lim} \approx 3.6 f^{2.50} (1 + z)^{2.03} E_{0.3}^{-3.0} \mGauss
\end{equation}
is the best lower limit on $B$ that can be derived even if the ARCADE best-fit radio background extends to $\nu \to \infty$.  For very high $B \gg B_{\rm lim}$ this means low energy primary electrons must be the source of the background.

The ARCADE background is marginally inconsistent with a secondary origin in most SFGs.  When $f = 1.0$, a secondary origin for the ARCADE excess is difficult to reconcile with the $\gamma$-ray background.  Eq.~\ref{eqn:BLim} rules out the intergalactic medium, clusters, and most galaxies at low redshifts.  Only the densest Ultraluminous Infrared Galaxies (ULIRGs) like Arp 220 and AGNs have the milliGauss magnetic fields needed \citep{Condon91,Torres04,Robishaw08}.  ULIRGs are among the \emph{brightest} (and therefore individually detected) galaxies, and cannot make up most of the ARCADE background \citep{Seiffert09}.  For a $z = 2$ population, eq.~\ref{eqn:BLim} gives us $B \ga 34 f^{2.5}\ \mGauss$ (corresponding to $3.1 f^{2.5}\ \GHz$), but we generally expect $B \la 20 - 40\ \mGauss$ for dynamical reasons \citep{Thompson06}.  For a source population at $z \ga 2$, eq.~\ref{eqn:BLim} no longer is the main restriction, and the minimum allowed $B$ decreases (eq.~\ref{eqn:BFromObs}).  Still, even at $z = 10$, a secondary origin for the ARCADE background at $\nu_{\rm max} = 3.4\ \GHz$ requires $B \ga 11 f^{-4.9}\ \mGauss$ (eq.~\ref{eqn:BFromObs}) in its sources.  Furthermore, there is little cumulative star formation at high $z$ \citep{Hopkins06}; since eq.~\ref{eqn:DiffSynchLimit} applies to each individual source population, the ARCADE sources would have to be extremely efficient at accelerating CR protons and contribute most of the $\gamma$-ray background.  Observations at 10 GHz can further constrain the possibility of a $z \approx 10$ source: at higher frequencies, there is more power in the radio background, but at higher energies, there is less power in the $\gamma$-ray background.  Eq.~\ref{eqn:FermiSynchLimit} demands a spectral turnover at high frequencies for the emission from secondary $e^{\pm}$.

The steep $f$ dependence means that uncertainties in the $\gamma$-ray background and kinematics weaken the constraints on $B$ considerably.  However, greater $f$ implies lower $B$, shifting the minimum allowed electron energy (0.15 $E_{0.3}$ GeV) to lower frequency; this relaxes the constraint in eq.~\ref{eqn:BLim}.  Even for $f = 2$, the 3.4 GHz detections requires milliGauss magnetic fields in the sources except at the highest redshifts.  Furthermore, the strong $f$ dependence works in reverse: if even half of the $\gamma$-ray background is not pionic, or not from the sources of the ARCADE excess, then the limits on $B$ strengthen by a factor $\sim 30$.  

Could the ARCADE excess be from primary electrons?  Any radio background from primaries can be accounted for if all of the protons escape.  However CR proton escape must be quite efficient; in the Milky Way, the luminosity of primary electrons is only $1 - 2\%$ that of CR protons at $\sim \GeV$ energies \citep{Schlickeiser02}.  Previous modeling indicates that secondaries are important in starbursts and perhaps the inner Galaxy, but unimportant for the low density outer Galaxy \citep{Porter08,Lacki10b}.  Recent work indicates that only $\sim 20\%$ of the Galactic GHz luminosity is from secondaries \citep{Strong10}; primaries can greatly enhance the radio background from low density SFGs.  However, galaxies have more gas at high $z$, making pionic losses more efficient.  Another possibility is that primary CR electron (but not proton) acceleration efficiency is much higher in some SFGs, producing more synchrotron.  \citet{Singal09} suggested AGNs provided such additional primaries.  IC upscattered starlight and bremsstrahlung in such galaxies would be a sign of these extra electrons.

\section{Conclusion}
\label{sec:Conclusion}
The observed $\gamma$-ray background limits the luminosity of pionic secondary $e^{\pm}$ in the Universe.  These secondary $e^{\pm}$ may be important in galaxy clusters and starburst galaxies.  We show that simple ratios can place bounds on the contribution of IC and synchrotron emission to the radio, X-ray, and $\gamma$-ray backgrounds from these secondary $e^{\pm}$.  With our given assumptions, the IC upscattered optical/UV light from secondaries contributes less than $f/4$ of the GeV $\gamma$-ray background for sources at $z = 10$ and smaller fractions at lower redshift and energies; upscattered FIR and CMB from secondaries is $\sim 2f\ \%$ or less of the 1 keV - 1 MeV X-ray background for sources at $z = 0$ and $\la 4\%$ at $z = 10$, although with uncertainties described below.  

We consider the ARCADE-measured radio background in light of these bounds.  Secondary $e^{\pm}$ are expected to dominate in starbursts that make up most of the star-formation at $z \ga 1$ \citep{Dole06,Caputi07,Magnelli09}.  The $\gamma$-ray background is marginally inconsistent with a secondary $e^{\pm}$ origin at $\sim 3\ \GHz$, unless the sources have milliGauss magnetic fields, although with considerable uncertainty.  However, we cannot rule out primary electrons in low density galaxies or other sources (where pionic losses are minimal) as the cause of the ARCADE measurement.

There are multiple uncertainties in these bounds.  First, we assumed the backgrounds all came from a source population with a single redshift, allowing considerable variation in the bounds as $z$ varies.  More detailed modelling of the effects of redshift evolution is needed.  Second, these constraints can be tightened by measuring the hadronic contribution to the $\gamma$-ray background \citep[e.g.,][]{Prodanovic04}; using the entire $\gamma$-ray background as done here may overestimate the other hadronic backgrounds.  Third, resolving out the contribution of each class of sources to the $\gamma$-ray background would tighten the limits on their contribution to the other backgrounds.  Note that this holds for the sources of the ARCADE excess specifically: even if most of the $\gamma$-ray background is star-formation, the sources of the ARCADE excess may contribute only a fraction of it.  Finally, future pionic neutrino background measurements above 100 GeV, such as with IceCube \citep[e.g.,][]{DeYoung09}, would help limit the IC and synchrotron backgrounds from the highest energy secondary $e^{\pm}$. 

\acknowledgments
I thank Todd Thompson and John Beacom for discussion and encouragement.  I also thank Eli Waxman for useful discussions, especially about the limits from equations~\ref{eqn:DiffSynchLimit} and~\ref{eqn:DiffICLimit}.  This work is funded in part by an Elizabeth Clay Howald Presidential Fellowship from OSU.

\end{document}